\documentclass[aps,prb,twocolumn,superscriptaddress,notitlepage]{revtex4-2}
\usepackage{graphicx}
\usepackage[caption=false]{subfig}
\usepackage{float}
\usepackage{natbib}
\usepackage{xcolor}
\usepackage{xr}
\usepackage{amsmath}
\usepackage{amssymb}
\usepackage{array}
\usepackage{wasysym}
\usepackage{color,soul}
\usepackage{braket}
\usepackage{verbatim}
\usepackage{hyperref}
\usepackage{gensymb}
\usepackage{url}
\usepackage{dirtytalk}

\def\smt{Sc$_6M$Te$_2$}
\def\sft{Sc$_6$FeTe$_2$}
\def\srt{Sc$_6$RuTe$_2$}
\def\sit{Sc$_6$IrTe$_2$}
\def\musr{$\mu$SR}
\def\brac{($M =~$Fe, Ru, Ir)}

\usepackage{filecontents}
\begin{document}

\title{Divergent Pressure Response of Superconductivity in \smt\ ($M$ = Fe, Ru and Ir)}

\author{J. N. Graham}
\email{jennifer.graham@psi.ch}
\affiliation{PSI Center for Neutron and Muon Sciences CNM, 5232 Villigen PSI, Switzerland}

\author{S.S. Islam}
%\thanks{These authors contributed equally to the experiments.}
\affiliation{PSI Center for Neutron and Muon Sciences CNM, 5232 Villigen PSI, Switzerland}

\author{K. Yuchi}
\affiliation{Institute for Solid State Physics, University of Tokyo, Kashiwa, Chiba 277-8581, Japan}

\author{P. Král}
\affiliation{PSI Center for Neutron and Muon Sciences CNM, 5232 Villigen PSI, Switzerland}

\author{O. Gerguri}
\affiliation{PSI Center for Neutron and Muon Sciences CNM, 5232 Villigen PSI, Switzerland}

\author{S. Huber}
\affiliation{PSI Center for Neutron and Muon Sciences CNM, 5232 Villigen PSI, Switzerland}
\affiliation{Physik-Institut, Universit\"{a}t Z\"{u}rich, Winterthurerstrasse 190, CH-8057 Z\"{u}rich, Switzerland}

\author{J.~Chang}
\affiliation{Physik-Institut, Universit\"{a}t Z\"{u}rich, Winterthurerstrasse 190, CH-8057 Z\"{u}rich, Switzerland}

\author{R. Khasanov}
\affiliation{PSI Center for Neutron and Muon Sciences CNM, 5232 Villigen PSI, Switzerland}

\author{Y. Okamoto}
\affiliation{Institute for Solid State Physics, University of Tokyo, Kashiwa, Chiba 277-8581, Japan}

\author{Z. Guguchia}
\email{zurab.guguchia@psi.ch} 
\affiliation{PSI Center for Neutron and Muon Sciences CNM, 5232 Villigen PSI, Switzerland}

\date{\today}

\begin{abstract}
Identifying and understanding non-BCS superconductivity remains a central challenge in condensed-matter physics. Here we focus on the \smt\  family ($M =d$-electron metal), which provides a unique platform of isostructural compounds exhibiting superconductivity across $3d$, $4d$, and $5d$ systems. Using hydrostatic pressure as an additional tuning parameter, muon-spin rotation (\musr) and AC susceptibility measurements uncover strongly contrasting pressure responses of superconductivity across the \smt\ series. The superconducting transition temperature, $T_\mathrm{C}$ decreases under pressure in the $3d$ Fe-based compound but increases for the $4d$ Ru- and $5d$ Ir-based systems, with the Ru compound showing the largest enhancement of nearly $50~\%$ within 2 GPa. The superfluid density exhibits similarly distinct pressure dependences, remaining nearly pressure independent for Fe while decreasing with increasing pressure for Ru and Ir. This suggests fundamentally different correlations between $T_\mathrm{C}$ and the superfluid density.
Together, these results indicate that superconductivity emerging from strongly correlated and spin–orbit–dominated regimes in \smt\ is likely governed by different microscopic mechanisms and offer a useful experimental basis for future microscopic theoretical studies.

%Together, these results reveal clear non-BCS characteristics of superconductivity in the \smt\ family and offer a useful experimental basis for future microscopic theoretical studies.

\end{abstract}
\maketitle
\section{Introduction}
The ability to design materials with specific properties is a compelling challenge in modern materials research. To enable such design, systematic studies across isostructural compounds are first required to understand how to manipulate physical characteristics with the chemistry toolkit. Electron correlations are widely regarded as a key ingredient for unconventional superconductivity and are realised in a broad class of correlated materials, including high-$T_{\rm c}$ cuprates \cite{doi:10.1126/science.237.4819.1133,keimer2015quantum,Uemura1,greene2020strange,rahman2015review,hayden2023charge}, Fe-based superconductors \cite{Fernandes_2022,mazin2010superconductivity,fernandes2014drives}, heavy-fermion systems \cite{RevModPhys.56.755}, Sr$_{2}$RuO$_{4}$ \cite{10.1063/1.1349611,mackenzie2003superconductivity}, kagome $A$V$_{3}$Sb$_{5}$ ($A$=K,Rb,Cs) compounds \cite{AV3Sb5_1,AV3Sb5_chiral3,kagome_vH3,mielke2022time,guguchia2023unconventional,graham2024depth,wang2023quantum}, and the nickelates \cite{wang2024experimental,nomura2022superconductivity,khasanov2025pressure}. Despite these numerous examples, it remains unclear which specific combinations of $d$-electron metals and crystal structures give rise to unconventional superconducting ground states. Spin–orbit coupling represents another central ingredient in quantum materials \cite{annurev:/content/journals/10.1146/annurev-conmatphys-020911-125138}, capable of stabilizing unconventional superconducting phases. Contributions from electron–lattice interactions \cite{Keller2005} have also been discussed. The ability to tune, via chemical substitution within an isostructural family and across different \textit{d}-electron metals, involves balancing  electron correlations, lattice effects, and spin–orbit coupling and therefore is a powerful route to systematically explore unconventional superconductivity. The case for such studies on $d$-electron superconductors was lacking until relatively recently with the discovery of the \smt\ \cite{Sc6MTe2} compounds.

The \smt\ compounds all adopt the $P\bar{6}2m$ structure without inversion symmetry and do not undergo a structural phase transition between $1.5~$K to $450~$K \cite{maggard2000sc6mte2, chen2002synthesis, Sc6MTe2, graham2025tailoring}. The structure is comprised of Sc atoms in a distorted trigonal prismatic environment surrounding a $d$-electron metal. These Sc$_6M$ clusters are then connected in one-dimensional chains along the $c$-axis, with each chain lying at the centre of a hexagonal Te net. This structure is robust to chemical substitution from $3d$ to $5d$ systems, and recently superconductivity was found in nearly all compositions \cite{Sc6MTe2}. As such, this structural consistency makes the \smt\ compounds an ideal family to study how superconductivity evolves across varying strengths of spin-orbit coupling and electron correlations of $d$-electron metals.

Initially, it may be surprising the family displays superconductivity at all, given that many $d$-electron metals are more usually associated with magnetism. In this regard, the chemical structure is not only important from a systematics standpoint, but also the large difference in electronegativity which means that the Sc atoms donate lots of electrons to the $M$ atoms, filling their outer shells and dramatically suppressing the magnetism \cite{PhysRevB.110.104505}. Furthermore, the lack of inversion symmetry allows for more unconventional Cooper pairing mechanisms to form beyond Bardeen-Cooper-Schrieffer (BCS) theory \cite{fischer2023superconductivity, sigrist2009introduction}, as found in other non-centrosymmetric compounds like CePt$_3$Si \cite{samokhin2004cept, bauer2007heavy} and La$_7$Ir$_3$ \cite{barker2015unconventional, li2018evidence, mayoh2021evidence}.
In the \smt\ family, superconductivity was found to be highest for the $3d$ systems (\sft $= 4.7~$K), with systematic variation according to their atomic number, but lower, around $2~$K for $4d$ and $5d$ systems \cite{Sc6MTe2}. The higher $T_\mathrm{C}$ of the Fe compound is thought to be due to its strong electron-phonon coupling, as demonstrated by low frequency vibrations associated with Rattling phonons as presented in an \textit{ab initio} study \cite{PhysRevB.110.104505}, higher than expected Sommerfeld coefficients \cite{Sc6MTe2}, and the specific heat capacity well exceeding the weak coupling expected from BCS superconductors \cite{Sc6MTe2, doi2025anisotropic}. Additionally, evidence for unconventional superconductivity in the Fe, Ru and Ir \smt\ compounds lies in their upper critical fields exceeding the Pauli limit \cite{Sc6MTe2}, and from muon spin rotation/relaxation (\musr) studies which found very dilute superfluid densities, exceptionally long London penetration depths ($> 1000~$nm) for \srt\ and \sit, and a slight paramagnetic shift of the internal field for \sit \cite{graham2025tailoring}. \musr\ has proposed all \smt ($M = ~$Fe, Ru, Ir) compounds have nodeless pairing symmetries with \sft\ having two gaps \cite{graham2025tailoring}, however NMR found a power law dependence for \srt\ indicating an anisotropic gap structure \cite{doi2025anisotropic}. Another essential feature of unconventional superconductivity is a competing or co-operative phase that may exist orders of magnitude above $T_\mathrm{C}$ in the normal state. The \musr\ study, in combination with neutron diffraction and magnetic susceptibility, found evidence that each \smt ($M = ~$Fe, Ru, Ir) compound has such a state, is likely magnetic in origin for \sft, and remains unknown for \srt\ and \sit, but the studies have excluded magnetic and structural origins \cite{graham2025tailoring}.

The delicate interplay of superconducting and normal states has been used as an advantage in leveraging superconducting properties in the past. For example, it was found in the kagome based $A$V$_3$Sb$_5$ ($A = ~$K, Rb, Cs) systems that following the complete suppression of charge order using hydrostatic pressure, superconductivity could be maximised and saturated at over $3$-$4$ times their ambient pressure values \cite{gupta2022microscopic, guguchia2023tunable, graham2025pressure, zheng2022emergent, neupert2022charge}. A similar effect has also been seen in LaRu$_3$Si$_2$ \cite{ma2025correlation, li2025superconducting}, $6$R-TaS$_2$ \cite{sazgari2025competing, liu2015tuning} and YBa$_2$Cu$_3$O$_7$ \cite{putzke2018charge, cyr2018sensitivity}.
However, the question still remains how the interplay of competing orders in the Sc$_6M$Te$_2$ series impacts the superconducting ground state, and to what effect spin-orbit coupling and electron correlations play in these matters. Therefore, in this article we present a highly systematic study using \musr\ and AC susceptibility under hydrostatic pressure to elucidate the pressure dependence of the \smt\ compounds. 
Our results reveal strikingly divergent behaviours of the superconducting transition temperature, $T_\mathrm{C}$, the superfluid density, and their mutual correlation across the \smt\ series. Among the compounds studied, the $4d$ system \srt\ exhibits the strongest sensitivity to applied pressure. These findings demonstrate that the pressure response of superconductivity is strongly governed by the choice of transition metal and the interplay between electronic correlations and spin–orbit coupling, providing important guidance for the development of a microscopic theory of superconductivity in Sc$_6M$Te$_2$.

\section{Superconducting state}
\begin{figure*}
    \centering
    \includegraphics[width=\textwidth]{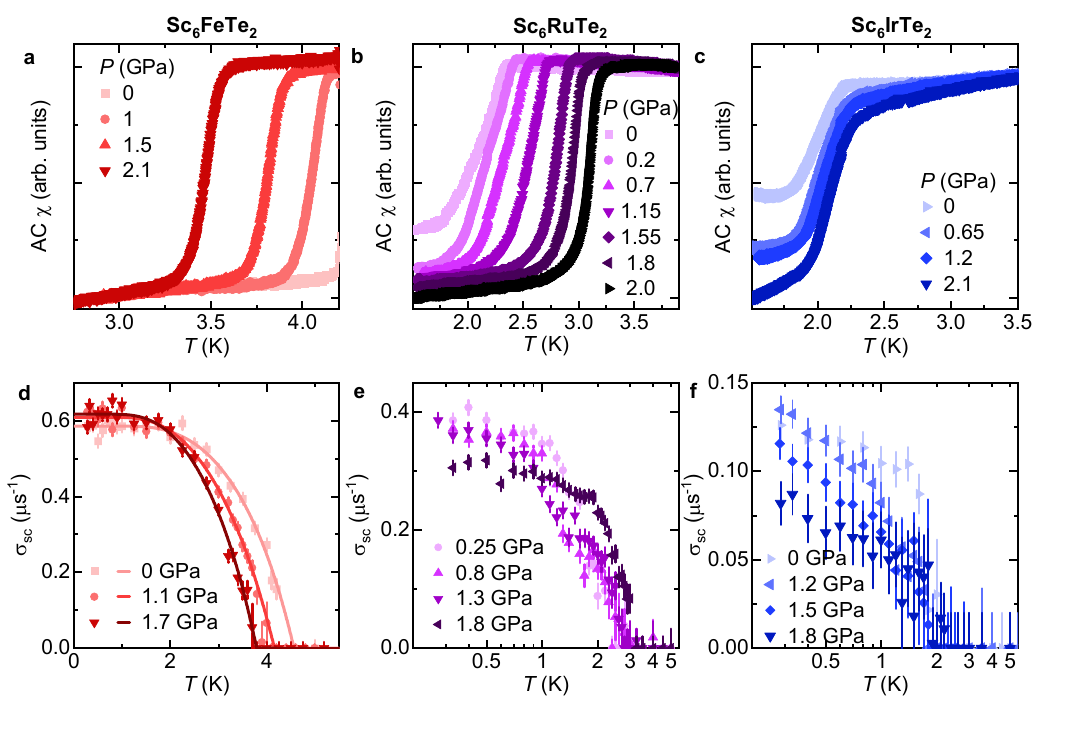}
    \caption{\textbf{Summary of AC susceptibility and \musr\ measurements under pressure on the \smt\ \brac\ compounds.} AC susceptibility measurements under various applied pressures up to $\sim 2.1~$GPa for \textbf{a} \sft, \textbf{b} \srt, and \textbf{c} \sit. \musr\ measurements under various applied pressures up to 1.8 GPa for \textbf{d} \sft, \textbf{e} \srt, and \textbf{f} \sit. Fits of a nodeless \textit{s}-wave superconducting gap structure are shown for \sft\ by the solid lines in \textbf{d}. \musr\ data are shown on a log-scale for \srt (\textbf{e}) and \sit (\textbf{f}) due to the limited temperature range. }
    \label{fig_1}
\end{figure*}
AC susceptibility and \musr\ measurements were performed on arc-melted  samples of \smt\ \brac\ under hydrostatic pressure up to $2.1~$GPa, as summarised in Fig. \ref{fig_1}. Firstly, we will examine the divergent behaviour of the critical temperature, $T_\mathrm{C}$ with pressure. Figure \ref{fig_1}a shows the response of \sft\ with pressure, where $T_\mathrm{C}$ decreases by about $1~$K between ambient and the highest applied pressure. It should be noted the full transition at $0~$GPa was not able to be fully measured due to technical limitations of the cryostat, however, this ambient pressure transition was verified for this sample via other methods (resistivity and DC susceptibility) \cite{Sc6MTe2, Sc6MTe2}. In contrast, \srt\ and \sit\ (Fig. \ref{fig_1}b and c, respectively) both have an overall increase in $T_\mathrm{C}$ with pressure, as well as a change in the magnitude of the diamagnetic response. Intriguingly, although \srt\ and \sit\ have nearly identical superconducting properties in ambient conditions \cite{graham2025tailoring}, they have very different sensitivities to pressure, with the $T_\mathrm{C}$ of \srt\ increasing by over $50~\%$ within $2~$GPa of applied pressure. 

To explore the microscopic properties of the superconductivity, \musr\ measurements under pressure were performed on each of the samples and are summarised in Fig. \ref{fig_1}d - f. The \sft\ sample has a clear suppression of $T_\mathrm{C}$, concomitant with the AC susceptibility, but the superfluid density stays nearly unchanged. An analysis of the superconducting gap structure was performed according to the supplemental material (SM), where each pressure was fit with a single nodeless $s$-wave model, and the results are summarised in Table \ref{tab_Fe}. This is different from the analysis at ambient pressure in Ref. \cite{graham2025tailoring}, where the \sft\ sample was best fit with a double nodeless gap, however, this discrepancy is likely due to the additional background from the pressure cell which makes resolving subtle changes, like a small secondary gap, challenging. Nevertheless, the analysis shows the London penetration depth is stable over the whole pressure range, but the size of the superconducting gap decreases by about $30~\%$.  

\begin{table}
    \centering
    \begin{tabular}{c|ccc}
        \hline
          P (GPa)&0&  1.1& 1.7\\
          \hline
 $\lambda^{-2} (T = 0~\mathrm{K}) (\mu\mathrm{m}^{-2})$& 5.44(5)& 5.66(5)&5.74(5)\\
          $\lambda (T > 0~\mathrm{K}) (\mathrm{nm})$&429&  420& 418\\
          $T_\mathrm{C}$ (K)&4.57(6)&  4.16(5)& 3.79(4)\\
 $\Delta$ (meV)& 0.98(3)& 0.82(2)&0.77(2)\\
 \hline
    \end{tabular}
    \caption{Summary of superconducting gap structure parameters under pressure for \sft. All the data were fit with a single nodeless $s$-wave model. $\lambda$ is the London penetration depth, $T_\mathrm{C}$ is the critical superconducting temperature and $\Delta$ is the size of the gap.}
    \label{tab_Fe}
\end{table}

The analysis and interpretation of the \srt\ and \sit\ compounds is different to \sft. Firstly, technically the \musr\ signal is much harder to isolate in these compounds due to the much weaker superfluid density in ambient conditions \cite{graham2025tailoring} which is further diluted from the additional background of the pressure cell, the reduction of $T_\mathrm{C}$ and the limited temperature range of the instrument ($T_\mathrm{min} = 300~$mK). Therefore, we offer only a qualitative discussion of the superconducting gap structure. To isolate the magnetic signal, we performed a background subtraction of the pressure cell as described in the SM. The critical temperature, $T_\mathrm{C}$, is difficult to precisely determine due to the scattering of points but has already been verified through AC susceptibility measurements, and the \musr\ results have a similar increase. As a result, we have put the \musr\ data on a log-scale to emphasise the evolution of the superfluid density at low temperatures. Both \srt\ and \sit\ appear to be consistent with fully gapped behaviour within the accessible temperature range. Intriguingly though, in contrast to \sft, both \srt\ and \sit\ have a reduction in their superfluid density with pressure. This is opposite from what was expected in the AC susceptibility measurements, where the magnitude of the curve increased with pressure. However, the AC magnitude is not geometry corrected and can be influenced by coil position and demagnetisation, therefore no direct comparison to superfluid density should be made.

\section{Normal state}

\begin{figure}
    \centering
    \includegraphics[width=0.5\textwidth]{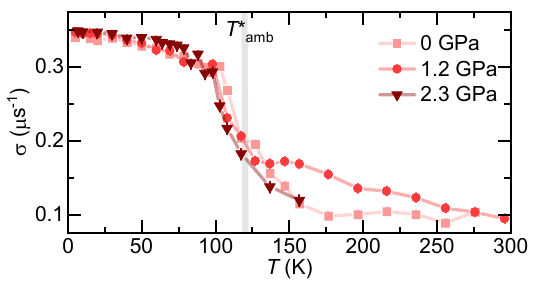}
    \caption{\textbf{Evolution of $T^*$ transition in the normal state of \sft.} The change in relaxation rate, indicating a phase transition ($T^*$), does not change with pressure. The feature is slightly broadened from the transition determined in ambient conditions ($T^*_\mathrm{amb}$ \cite{graham2025tailoring}) which is likely due to the pressure cell subtraction.}
    \label{fig_normal}
\end{figure}

Since other studies have shown that suppressing the normal state via pressure can increase the superconducting properties or vice versa, we have explored how the normal state of the \smt\ \brac\ compounds responds to pressure. As a reminder, a previous \musr\ study identified large increases/broadening in the normal state relaxation at $T^* = 120~$K for \sft, $260~$K for \srt\ and $390~$K for \sit, which were associated with phase transitions in the normal state \cite{graham2025tailoring}. \sft\ is the only compound we can measure the normal state of under pressure due to the \sit\ transition being outside the temperature range of the cryostat ($5$ - $300~$K), and the non-linear dependence of the pressure cell contribution at high temperatures makes interpreting data near room temperature challenging (\srt). The evolution of the normal state with pressure for \sft\ is shown in Fig. \ref{fig_normal}. The data were fit with a Gaussian Kubo-Toyabe function multiplied by a simple exponential (Eq. \ref{Eq_GKT} in SM). This is the same model used for the analysis of the ambient pressure data in Ref. \cite{graham2025tailoring}. The ambient pressure measurement matches well the results presented in Ref. \cite{graham2025tailoring}, though the transition is slightly broadened. With the application of pressure, no significant change in the normal state transition is seen, despite the clear decrease in $T_\mathrm{C}$ over the same pressure range.

\begin{figure*}
    \centering
    \includegraphics[width=\textwidth]{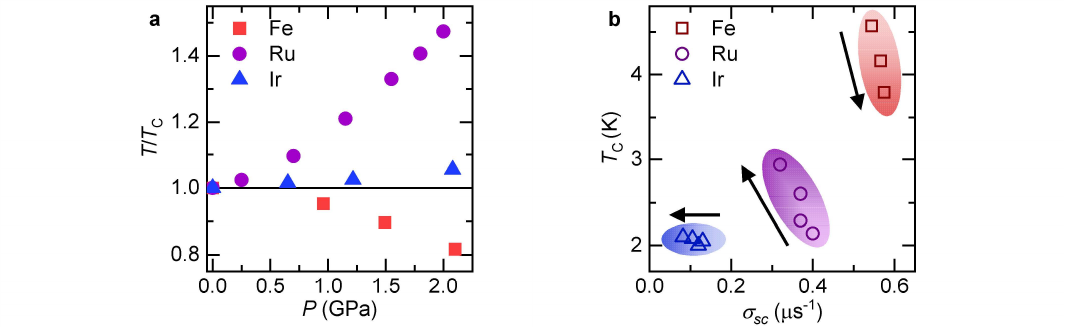}
    \caption{\textbf{Summary of pressure response of the \smt \brac\ compounds. a} Relative change of $T_\mathrm{C}$ of each \smt \brac\ compound with pressure. $T_\mathrm{C}$ is normalised to the ambient pressure value to put all compounds on the same scale. \srt\ has overall the largest change with pressure. \textbf{b} Relationship between $T_\mathrm{C}$ and superfluid density, $\sigma_\mathrm{sc}$ where all three compounds show contrasting responses with pressure. Arrows and coloured ovals (light to dark) denote increasing hydrostatic pressure.}
    \label{fig_summary}
\end{figure*}

\section{Discussion}

Taken together, this complementary set of \musr\ and AC susceptibility measurements demonstrates that the pressure response of the superconducting properties in the family \smt\ \brac\ depends strongly on the choice of transition metal, reflecting a crossover from dominant electronic correlations in the Fe-based compound to strong spin–orbit coupling in the Ir-based system. The main findings of this study can be summarised as follows: 

(1) The critical temperature, $T_\mathrm{C}$, has divergent behaviour with pressure as summarised in Fig. \ref{fig_summary}a. Figure \ref{fig_summary}a shows the change of $T_\mathrm{C}$ relative to its ambient pressure value. Here, it can clearly be seen that $T_\mathrm{C}$ for the Fe sample decreases, whereas $T_\mathrm{C}$ for the Ru and Ir samples increase with pressure. Overall \srt\ is the most sensitive to pressure, increasing by $50~\%$ within $2~$GPa, followed by an overall change of $20~\%$ for \sft\ and less than $5~\%$ for \sit. The enhancement of $T_\mathrm{C}$ observed in both the Ir- and Ru-based compounds is particularly noteworthy, as conventional BCS superconductors typically exhibit a suppression of $T_\mathrm{C}$ under pressure. This behaviour suggests that applying higher pressures may further enhance $T_\mathrm{C}$, naturally raising the question of whether the transition temperature will continue to increase, eventually saturate, or evolve into a dome-shaped pressure dependence. These findings strongly motivate future experiments extending to significantly higher pressures than those explored in the present study.

(2) Besides $T_\mathrm{C}$, the superfluid density is a fundamental superconducting parameter that also exhibits distinctly different pressure responses across the three compounds (Fig. \ref{fig_summary}b). In \sft\, the superfluid density remains essentially unchanged under pressure despite a clear suppression of $T_\mathrm{C}$. In contrast, for both \srt\ and \sit\, the superfluid density decreases with increasing pressure while $T_\mathrm{C}$ is simultaneously enhanced, with the Ru-based compound again displaying the strongest pressure sensitivity. These contrasting trends demonstrate that the correlation between $T_\mathrm{C}$ and  superfluid density differs fundamentally among the three systems.
The relationship between $T_\mathrm{C}$ and the superfluid density is widely regarded as a hallmark of unconventional superconductivity, and our results clearly place the \smt\ family within this class. Remarkably, however, the nature of this correlation depends strongly on whether superconductivity is governed by $3d$, $4d$, or $5d$ electrons. In comparison to other superconductors, the cuprates have the well-known Uemura \cite{Uemura1} relation which establishes a linear scaling between $T_\mathrm{C}$ and the superfluid density in the underdoped regime, while near optimal doping the superfluid density increases with little change in $T_\mathrm{C}$. Upon further overdoping, both quantities decrease, giving rise to the so-called “boomerang effect”\cite{PhysRevLett.71.1764}. This boomerang behaviour has long challenged simple theoretical descriptions of high-$T_\mathrm{C}$ superconductivity and has been attributed to several mechanisms, including strong phase fluctuations, intrinsic disorder or inhomogeneity, and competition or coexistence with other ordered states such as charge-density waves. In the case of \smt\, the three distinct $T_\mathrm{C}$–superfluid density correlations qualitatively resembles the behaviour observed in the optimally doped and overdoped regimes of cuprates \cite{PhysRevLett.71.1764}, highlighting a complex interplay between electronic correlations and spin–orbit coupling that evolves systematically from $3d$ to $5d$ systems. This parallel between the cuprates and \smt\ is highly encouraging and provides a valuable framework for theory, underscoring the importance of a microscopic understanding of phase fluctuations, quantum criticality, and anharmonic electron–lattice interactions in shaping the relationship between  $T_\mathrm{C}$ and the superfluid density.

(3) The normal state transition, $T^*$ for \sft\ does not change with pressure (Fig. \ref{fig_normal}). This indicates that the observed changes in the superconducting properties are not directly linked to modifications of the normal-state behaviour. Instead, it suggests that distinct electronic states are involved in the formation of superconductivity and the normal-state transition. This behaviour differs from the cooperative or competitive interplay commonly observed between superconductivity and normal-state transitions, such as charge order or magnetism, in many other superconducting systems such as LaRu$_3$Si$_2$ \cite{ma2025correlation, li2025superconducting}, $6$R-TaS$_2$ \cite{sazgari2025competing, liu2015tuning}, La$_{2-x}$Ba$_x$CuO$_4$ \cite{guguchia2020using} and YBa$_2$Cu$_3$O$_7$ \cite{putzke2018charge, cyr2018sensitivity}. The pressure effect between the normal state properties of \srt\ and \sit\ and superconductivity could not be explored in this study due to technical limitations of the cryostat and the pressure cell setup. However, earlier work found that the origin of these $T^*$ transitions were different for \sft\ and Sc$_6$(Ru/Ir)Te$_2$ (magnetic and non-magnetic/structural, respectively). Therefore, one area that should be explored is firstly what is the origin of the $T^*$ transition for Sc$_6$(Ru/Ir)Te$_2$, and then is this why the compounds have contrasting responses to pressure in the superconducting state? Measurements such as ARPES, high intensity X-ray diffraction or RIXS are suitable techniques to explore the first question, once single-crystal samples of the \smt\ family are available.

\section{Conclusion and outlook}
In conclusion, these results reveal intriguing divergent behaviours of how pressure and chemical tuning can impact the physics of three new $d$-electron based superconductors \smt\ ($M$=Fe, Ir, Ru). In the first instance, it is notable that the superconductor with the optimal  $T_\mathrm{C}$ among three compounds in ambient conditions, \sft, is destroyed with the application of pressure.
In contrast, \srt\ and \sit\ exhibit superconducting transition temperatures that are approximately a factor of two lower than that of \sft\ under ambient conditions, but their superconductivity is enhanced upon the application of pressure. Most curiously, is that the sensitivity of these two systems to pressure is very different, despite their nearly identical behaviour in ambient conditions. Ultimately, the physics of these systems appears to be governed by the interplay between electron correlations and spin–orbit coupling, yet how these effects manifest in $d$-electron superconductors remains an open question. This motivates further theoretical work to elucidate how $d$-electron metallic states set the balance between electron correlations and spin–orbit coupling, how these interactions evolve under pressure, and the role of electron–lattice coupling. Experimentally, future studies should focus on the influence of the normal state on superconductivity in the \smt compounds and on extending the accessible pressure range to track the evolution of $T_\mathrm{C}$. Understanding the distinct responses of each system will be essential for developing a unifying framework for non-BCS superconductivity in $d$-electron materials.

\section{Methods}
\subsection*{AC Susceptibility}
AC susceptibility data were collected using an in-house glass cryostat system at the Paul Scherrer Institute, Villigen, Switzerland. Arc melted pellet samples were cut up to fit within the pressure chamber of a double wall MP35N/MP35N piston cell. A small amount of Daphne 7373 oil was used as the pressure medium. This cryostat setup has a maximum temperature of $4.2~$K, therefore the full transition of \sft\ at $0~$GPa was not able to be fully measured, however, this ambient pressure transition was verified for this sample via other methods (resistivity and DC susceptibility) \cite{Sc6MTe2}. The pressure was determined by tracking the superconducting transition of indium against a reference indium outside the pressure cell, both of which were removed from the susceptibility curves for clarity. The absolute value of the susceptibility of each sample was artificially altered to account for differences in the coil position, but the magnitude of each curve was not corrected. This method was also used to determine the applied pressure for the \musr\ experiments where the sample was loaded into a MP35N/CuBe pressure cell.

\subsection*{\musr\ Experiments}

A muon spin rotation/relaxation (\musr) experiment consists of a beam of nearly $100~\%$ spin polarised muons, $\mu^+$, which are implanted into the sample, one muon at a time. These positively charged muons stop, due to thermal stabilisation, at interstitial lattice sites in the crystal. The muon then precesses at a frequency proportional to the local internal magnetic field, $B_\mu$, before decaying into a positron which is preferentially emitted in the direction of the muon spin, which is then detected. 

\subsection*{Experimental details}

Transverse-field (TF) and zero-field (ZF) \musr\ experiments under pressure were conducted at the high pressure spectrometer, GPD \cite{khasanov2022perspective} at the Swiss Muon Source (S$\mu$S), Paul Scherrer Institute, Villigen, Switzerland. Arc melted pellet samples were cut up/crushed to fit within the pressure chamber of a double wall CuBe/MP35N piston cell. The double wall cell is specifically designed for high-pressure measurements and can reach applied pressures of 2.3 GPa at room temperature. The CuBe/MP35N alloy was chosen because it has a lower background for \musr\ experiments, which was required due to the very dilute superfluid signal. Daphne 7373 oil was used as a pressure medium. For the measurements in the superconducting state, a Variox He-4 cryostat with a He-3 insert was used to control the temperature, allowing temperatures of $300~$mK. A TF of $30~$mT was applied to \sft, and $20~$mT applied to \srt\ and \sit. These TFs were chosen to match the ambient pressure TFs applied in Ref. \cite{graham2025tailoring}. The field was applied at 10 K, and the samples cooled in field, to access the vortex state. For the ZF measurements in the normal state, a Janis cryostat, with a temperature range of $5$ - $300~$K, was used to control the temperature. Measurements of the pressure cell background were made by lowering the muon momentum to $98~$MeV/c, so that the majority of the muons stop within the cell. These measurements were only performed under an applied TF of $20~$mT between $300~$mK and $5~$K. 

\subsection*{Data reduction and analysis}
\begin{figure}
    \centering
    \includegraphics[width=0.5\textwidth]{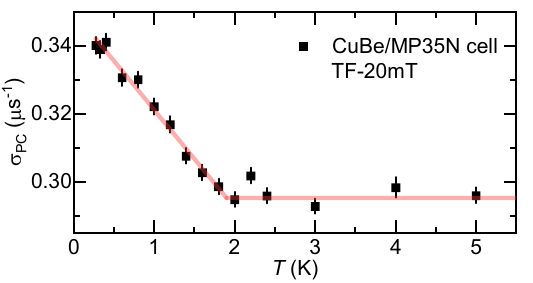}
    \caption{The temperature dependence of the relaxation rate of the CuBe/MP35N pressure cell with the sample inside is shown. The measurements were performed in a $20~$mT applied transverse field at a muon momentum of 98 MeV/c, for which the majority of muons stop in the pressure cell. The relaxation rate exhibits a non-linear temperature dependence, with only a very weak increase below 2 K attributable to the influence of the superconducting sample.}
    \label{fig_cell_bkg}
\end{figure}
The superfluid density can be accurately measured by the \musr\ technique from the evolution of the muon spin depolarisation rate, $\sigma_\mathrm{tot}$ as a function of temperature. This depolarisation rate is proportional to the width of the internal field experienced by the muon, which for a superconductor is comprised of two components, a superconducting, $\sigma_\mathrm{sc}$ and nuclear magnetic dipolar, $\sigma_\mathrm{nm}$ contribution ($\sigma_\mathrm{tot}^2 = \sigma_\mathrm{sc}^2 + \sigma_\mathrm{nm}^2)$. For measurements under hydrostatic pressure there is an additional contribution from the muons that stop within the pressure cell, and can thus be considered as an additional background. This is particularly problematic for \srt\ and \sit, which from our \musr\ measurements in ambient conditions had maximal $\sigma_\mathrm{sc}$ values of $0.15~\mu\mathrm{s}^{-1}$ and $0.1~\mu\mathrm{s}^{-1}$, respectively \cite{graham2025tailoring}. To address this issue, we performed an additional measurement at $20~$mT using a reduced muon momentum, such that the majority of muons stop in the pressure cell. The sample remains inside the cell and therefore still influences the pressure-cell signal, which is precisely what we aim to probe, as this configuration allows us to directly characterize the pressure-cell background under the same experimental conditions as the sample measurements. The pressure cell was fit with the following function:
\begin{equation}
    A_\mathrm{PC}(t) = A_\mathrm{PC}\mathrm{exp}\left[-\frac{1}{2}(\sigma_\mathrm{PC} t^2)\right]\mathrm{cos}(\gamma_\mu B_\mathrm{PC}t + \varphi_\mathrm{PC})
    \label{Eq_PC}
\end{equation}
where $A_\mathrm{PC}$ is the initial asymmetry of the pressure cell, $\sigma_\mathrm{PC}$ is the depolarisation rate of the pressure cell, $\gamma_\mu$ is the gyromagnetic ratio of the muon, $B_\mathrm{PC}$ is the internal field of the pressure cell and $\varphi_\mathrm{PC}$ is the phase shift of the pressure cell. As shown in Fig. \ref{fig_cell_bkg}, the pressure cell exhibits a very weak increase in the relaxation rate below 2 K, which arises from the influence of the superconducting sample. Although this increase is small, given that the superfluid responses of both the Ir- and Ru-based samples are weak, we interpolate this non-linear increase (red line) and use it as the background contribution in our data reduction. The background was subtracted in quadrature from $\sigma_\mathrm{tot}$ for both \srt\ and \sit.

The \sft, \srt\ and \sit\ superconducting data were fit with the following function:
\begin{equation}
     A_\mathrm{TF}(t) = A_\mathrm{S}\mathrm{exp}\left[-\frac{1}{2}(\sigma_\mathrm{S} t^2)\right]\mathrm{cos}(\gamma_\mu B_\mathrm{int,S}t + \varphi_\mathrm{S}) + A_\mathrm{PC}(t)
\end{equation}
where $A_\mathrm{S}$ is the initial asymmetry of the sample, $\sigma_\mathrm{S}$ is the depolarisation rate of the sample, $B_\mathrm{int,S}$ is the internal field of the sample, $\varphi_\mathrm{S}$ is the phase shift of the sample and $A_\mathrm{PC}(t)$ is the pressure cell contribution, as defined in Eq. \ref{Eq_PC}. 

Following the background subtraction, $\sigma_\mathrm{sc}$ can be estimated by subtracting in quadrature $\sigma_\mathrm{nm}$ (assumed to be constant above $T_\mathrm{C}$) from the corrected $\sigma_\mathrm{tot}$. $\sigma_\mathrm{sc}$ for each pressure is shown in Fig. \ref{fig_1}d - f. 

\begin{figure}
    \centering
    \includegraphics[width=0.5\textwidth]{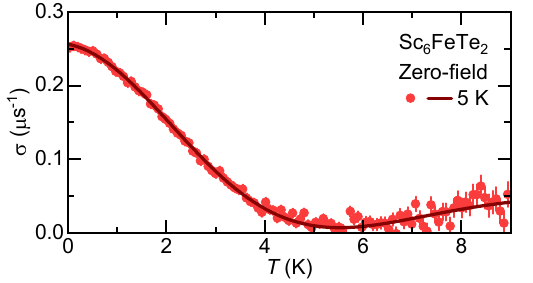}
    \caption{Example fit of zero-field \musr\ data in the normal state of \sft\ at $5~$K. Data were fit with Eq. \ref{Eq_GKT}.}
    \label{Fig_GKT}
\end{figure}

The normal state zero-field data of \sft\ were fit with the following Gaussian Kubo-Toyabe depolarisation function multiplied by an exponential term
\begin{equation}
    A^\mathrm{GKT}_\mathrm{ZF}(t) = \sum^n_i A\left[\left(\frac{1}{3} + \frac{2}{3}(1-\sigma^2_it^2)\mathrm{exp}\left[-\frac{\sigma^2_it^2}{2}\right]\right)\mathrm{exp}(-\Gamma_i t)\right]
    \label{Eq_GKT}
\end{equation}
where $\sigma_i$ and $\Gamma_i$ are the muon relaxation rates. For these measurements there are two $i$ components, a pressure cell and sample contribution. This formula is consistent with the functional form used in Ref. \cite{graham2025tailoring} for the ambient pressure data. An example fit of the data at $5~$K is shown in Fig. \ref{Fig_GKT}.

\subsection*{Superconducting gap structure analysis}
To perform a quantitative analysis of \musr\ data and determine the superconducting gap structure, the superconducting muon spin depolarisation rate, $\sigma_\mathrm{sc}(T)$ in the presence of a perfect triangular vortex lattice is first related to the London penetration depth, $\lambda(T)$ by the following equation \cite{London_muSR, London_muSR2}:
\begin{equation}
    \frac{\sigma_\mathrm{sc}(T)}{\gamma_\mu} = 0.06091\frac{\Phi_0}{\lambda^2(T)}
\end{equation}
where $\Phi_0 = 2.068 \times 10^{-15}~$Wb is the magnetic flux quantum. This equation is only applicable when the separation between the vortices is larger than $\lambda$. In this particular case, as per the London model, $\sigma_\mathrm{sc}$ becomes field independent. By analysing the temperature dependence of the magnetic penetration depth, within the local London approximation, a direct association with the superconducting gap symmetry can be made \cite{musrfit}:

\begin{equation}
    \frac{\lambda^{-2}(T, \Delta_{0,i})}{\lambda^{-2}(0, \Delta_{0,i})} = 1 + \frac{1}{\pi} \int_0^{2\pi} \int_{\Delta(T, \phi)}^\infty \left(\frac{\delta f}{\delta E} \right) \frac{EdEd\varphi}{\sqrt{E^2-\Delta_i(T,\varphi)^2}}
\end{equation}
where $f = [1+\mathrm{exp}(E/k_BT)]^{-1}$ is the Fermi function, $\varphi$ is the angle along the Fermi surface, and $\Delta_i(T, \varphi) = \Delta_{0,i}\Gamma(T/T_\mathrm{C})g(\varphi)$ ($\Delta_{0,i}$ is the maximum gap value at $T = 0$). The temperature dependence of the gap is approximated by the expression, $\Gamma(T/T_\mathrm{C} = \mathrm{tanh}{1.82[1.018(T_\mathrm{C}/T - 1)]^{0.51}}$ \cite{sc_gap}, whilst $g(\varphi)$ describes the angular dependence of the new gap and is replaced by $1$ for an $s$-wave gap, $[1+a\mathrm{cos}(4\varphi)/(1 + a)]$ for an anisotropic $s$-wave gap, and $|\mathrm{cos}(2\varphi)|$ for a $d$-wave gap \cite{sc_gap2}.\\ 

\bibliography{References.bib}

\section{Acknowledgments}~
Z.G. acknowledges support from the Swiss National Science Foundation (SNSF) through SNSF Starting Grant (No. TMSGI2${\_}$211750). Y.O. and K.Y. acknowledge support from the Japan Science and Technology Agency through JST-ASPIRE (No. JPMJAP2314).\\

\section{Author contributions}~
Z.G. conceived and supervised the project. Sample growth: K.Y. and Y.O.. High pressure AC susceptibility experiments: J.N.G., S.S.I., and Z.G.. High pressure $\mu$SR experiments, the corresponding analysis and discussions: J.N.G., S.S.I., P.K., O.G.,  R.K., and Z.G.. Figure development and writing of the paper: J.N.G. and Z.G. All authors discussed the results, interpretation, and conclusion.\\ 

\section*{Data availability}
The data that support the findings of this study are available from the corresponding authors upon request.\\

\section*{Conflict of Interest}
The authors declare no financial/commercial conflict of interest.\\

\end{document}